\documentclass[final,1p,times]{elsarticle}
\usepackage{graphics}
\usepackage{amssymb}
\usepackage{bm}
\begin{document}
\begin{frontmatter}

\title{ Exploring the QGP phase  above the deconfinement temperature in $pp$ and $A-A$ collisions at LHC energies}
\author[a,b]{Aditya Nath Mishra}, 
\affiliation[a]{organization={Department of Physics, University Institute of Sciences},
            addressline={Chandigarh University, Mohali- 140413}, 
            state={Punjab},
            country={India}}
          \affiliation[b]{organization={University Centre For Research $\&$ Development (UCRD)},
            addressline={Chandigarh University, Mohali- 140413},
            state={Punjab},
            country={India}} 
 \author  [c]{Guy Pai{\'c}},       
 \affiliation[c]{organization={Instituto de Ciencias Nucleares,
            Universidad Nacional Autonoma de Mexico},
             addressline={Apartado Postal 70-543},
            postcode={04510},
            country={Mexico }}
          \author  [d]{C. Pajares},
          \affiliation[d]{organization={Departamento de Fisica de Particulas, Universidale de Santiago de Compostela and Instituto Galego de Fisica de Atlas Enerxias(IGFAE)},
          addressline={15782 Santiago, de Compostela},
          country={Spain}}
        \author  [e]{R. P. Scharenberg},
        \affiliation[e]{organization={ Department of Physics and Astronomy, Purdue University},
          addressline={ West Lafayette, IN-47907},
          country={USA}}          
        \author  [e]{B. K. Srivastava\corref{cor1}}
        \cortext[cor1]{Corresponding author}
        \ead{brijesh@purdue.edu}

\date{\today}
\begin{abstract}
  In the present work we have analyzed the transverse momentum spectra  of charged particles in high multiplicity ${\it pp}$ collisions at LHC energies 
$\sqrt s = $ 5.02 and 13  TeV published by the ALICE Collaboration using the Color String Percolation Model (CSPM).
For heavy ions  Pb-Pb at $\sqrt {s_{NN}} =$ 2.76 and 5.02 TeV along with Xe-Xe at $\sqrt {s_{NN}} = $ 5.44 TeV have been analyzed.
The initial temperature is extracted both in low and high multiplicity events in ${\it pp}$ collisions. For $A-A$ collisions the temperature is obtained as a function of centrality.
 A universal scaling in the temperature from $pp$ and $A-A$ collisions is obtained when multiplicity is scaled by the transverse interaction area.
 From the measured energy density $\bm \varepsilon$ and the temperature the dimensionless quantity $\bm \varepsilon/T^{4}$ is obtained.
 Our results for Pb-Pb and Xe-Xe collisions show a sharp increase in $\bm \varepsilon/T^{4}$ above T $\sim$ 210 MeV and reaching the ideal gas of quarks and gluons value of $\bm \varepsilon/T^{4} \sim$ 16 at temperature $\sim $ 230 MeV. In case of $pp$ collisions only $\bm \varepsilon/T^{4} \sim $ 10 corresponding to  $\sim$ 30 degrees of freedom. 
 \end{abstract}
 \begin{keyword}
   Quark gluon plasma \sep Heavy-ion collisions
 \PACS{25.75.-q, 25.75.Gz, 25.75.Nq, 12.38.Mh}
 \end{keyword}

\end{frontmatter}
 
\section{Introduction}
The Quantum Chromodynamics (QCD) phase diagram is closely related to the history of the universe and can be probed by heavy ion collisions. Of particular interest in the heavy ion collision experiments are the details of the deconfinement and chiral transitions which determine the QCD phase diagram.  One of the main challenges of the field is to simultaneously determine the initial temperature and the energy density of the matter produced in a collision and hence the number of thermodynamic degrees of freedom (DOF) \cite{busza}.

Several interesting features related to the  $QGP$ formation e.g., long range rapidity correlations, the so called ``ridge'', elliptic flow and strangeness enhancement seen in heavy ion collisions are also observed in  high multiplicity $pp$ collisions at LHC energies \cite{CMS1,CMS2,CMS3,ATLAS,ALICE}.

The objective of the present work is to extract the thermodynamic properties like initial temperature and the degrees of freedom (DOF) in $pp$, Pb-Pb, and Xe-Xe collisions at LHC energies by analyzing the published ALICE data on the transverse momentum spectra of charged hadrons \cite{alicepp,alicepp7,alicepb,alicexe} using the framework of clustering of color sources (CSPM) \cite{review15,universe}. This requires the measurement of the initial thermalized temperature and the energy density at time $\sim$ 1 fm/c of the hot matter produced in these high energy hadron-hadron and nucleon-nucleon collisions. A letter was published using the limited sets of data \cite{epja21}. 

This approach has been successfully used to describe the initial stages in the soft region in high energy nucleus-nucleus and nucleon-nucleon collisions
\cite{eos,eos2,IS2013,eos3,cpod13,pp19,cunq,andres,iit1,iit2,iit3}. The CSPM is in fact different from the hydrodynamics picture and is more in line with other studies where the interaction among strings \cite{bierlich,bierlich2,ortiz} or the domain color structure \cite{muller,lappi} is taken into account.
Lattice Quantum Chromo Dynamics simulations (LQCD) indicate that the non-perturbative region of hot QCD matter extends up to temperature of 400 MeV \cite{latthigh}, well above the universal hadronization temperature \cite{bec1}.

The paper is organized as follows: section 2 describes the phenomenology of the color string percolation model. The measurement of color suppression factor $F(\xi)$ and its relation to temperature are presented in sections 3 and 4. Sections 5 to 8 describe thermalization, energy density, and degrees of freedom.      

\section{Clustering of color sources}
Multiparticle production is currently described in terms of color strings stretched between the projectile and the target, which decay into new strings through color neutral $q-\bar{q}$ pairs production and subsequently hadronize to  produce the observed hadrons \cite{review15,pajares1,pajares2}. Color strings may be viewed as small areas in the transverse plane filled with color field created by colliding partons. In terms of gluon color field they can be considered as the color flux tubes stretched between the colliding partons. The mechanism of particle creation is the Schwinger $QED_{2}$ mechanism and is due to the color string breaking \cite{schw,wong}.

 The initial colliding quarks and anti-quarks interact to form a large number of color strings. The non-perturbative Schwinger particle creating mechanism in quantum electrodynamics $QED_{2}$, with massless fermions, was derived in an exact gauge invariant calculation \cite{schw}. $QED_{2}$ contains a single space and time coordinate. Confinement, charge screening, asymptotic freedom and the existence of a neutral bound state boson in  $QED_{2}$ closely models $QCD$. When string color fields are present, the Schwinger $QED_{2}$ string breaking mechanism lifts color neutral $q\bar{q}$ pairs from vacuum \cite{wong}. 
String breaking proceeds in an iterative way until they come to objects with masses comparable to hadron masses, which can be identified with observable hadrons by combining the produced flavor with statistical weights \cite{schw,wong}. The Schwinger mechanism has also been used in the decay of color flux tubes produced by the quark-gluon plasma for modeling the initial  stages in heavy ion collisions \cite{prc1,prc2}. 

\begin{figure}[thbp]
\centering        
\vspace*{-0.2cm}
\resizebox{0.55\textwidth}{!}{
  \includegraphics{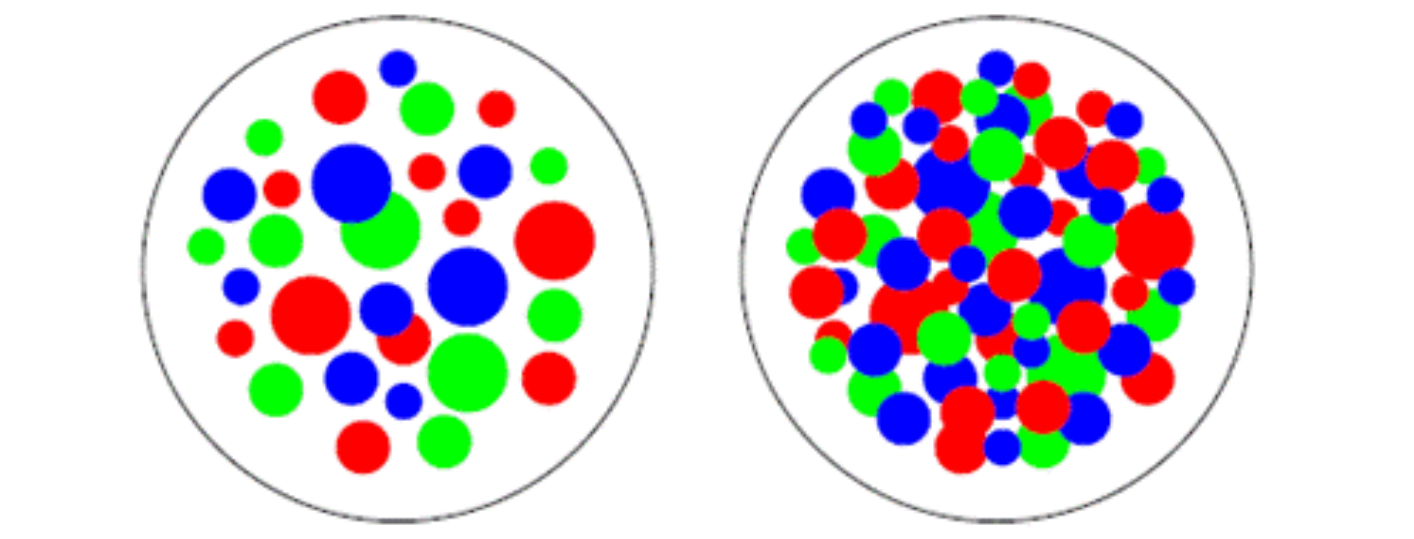}
}
\caption{ Partonic cluster structure in the transverse collision plane at low (left) and high (right) parton density \cite{satzbook}.}
\label{disk}
\end{figure}

With growing energy and size of the colliding system, the number of strings grows, and they start to overlap, forming clusters, in the transverse plane very much similar to disks in two dimensional percolation theory as shown in 
Fig.~\ref{disk} \cite{inch,satzbook}. At a certain critical density, a macroscopic cluster appears that marks the percolation phase transition. The interaction between strings occurs when they overlap and the general result, due to the SU(3) random summation of charges, is a reduction in multiplicity and an increase in the string tension, hence an increase in the average transverse momentum squared, $\langle p_{T}^{2} \rangle$. This is the Color String percolation Model (CSPM) \cite{pajares1,pajares2}. 
 We assume that a cluster of ${\it n}$ strings that occupies an area of $S_{n}$ behaves as a single color source with a higher color field $\vec{Q_{n}}$ corresponding to the vectorial sum of the color charges of each individual string $\vec{Q_{1}}$. The resulting color field covers the area of the cluster. As $\vec{Q_{n}} = \sum_{1}^{n}\vec{Q_{1}}$, and the individual string colors may be oriented in an arbitrary manner with respective to each other, the average $\vec{Q_{1i}}\vec{Q_{1j}}$ is zero, and $\vec{Q_{n}^2} = n \vec{Q_{1}^2} $. This results in the suppression of the multiplicity and the enhancement of $\langle p_{T}^{2}\rangle$.  
\\
\\
\\
Knowing the color charge $\vec{Q_{n}}$ one can obtain the multiplicity $\mu_{n}$ and the mean transverse momentum squared $\langle p_{T}^{2} \rangle_{n}$ of the particles produced by a cluster of $\it n $ strings \cite{pajares2}
\begin{equation}
\mu_{n} = \sqrt {\frac {n S_{n}}{S_{1}}}\mu_{1};\hspace{5mm}
\langle p_{T}^{2} \rangle_{n} = \sqrt {\frac {n S_{1}}{S_{n}}} {\langle p_{T}^{2} \rangle_{1}}
\label{mu1pt1}
\end{equation} 
where $\mu_{1}$ and $\langle p_{T}^{2}\rangle_{1}$ are the mean multiplicity and $\langle p_{T}^{2} \rangle$ of particles produced from a single string with a transverse area $S_{1} = \pi r_{0}^2$, where $r_{0}$ is the string radius. For strings  just touching each other $S_{n} = n S_{1}$, and $\mu_{n} = n \mu_{1}$, $\langle p_{T}^{2}\rangle_{n}= \langle p_{T}^{2}\rangle_{1}$. When strings fully overlap, $S_{n} = S_{1}$  and therefore 
$\mu_{n} = \sqrt{n} \mu_{1}$ and $\langle p_{T}^{2}\rangle_{n}= \sqrt{n} \langle p_{T}^{2}\rangle_{1}$, so that the multiplicity is maximally suppressed and the $\langle p_{T}^{2}\rangle_{n}$ is maximally enhanced. This implies a simple relation between the multiplicity and transverse momentum $\mu_{n}\langle p_{T}^{2}\rangle_{n}=n\mu_{1}\langle p_{T}^{2}\rangle_{1}$, which means conservation of the total transverse momentum produced.
\begin{figure}[thbp]
\centering        
\vspace*{-0.2cm}
\includegraphics[width=0.6\textwidth,height=3.0in]{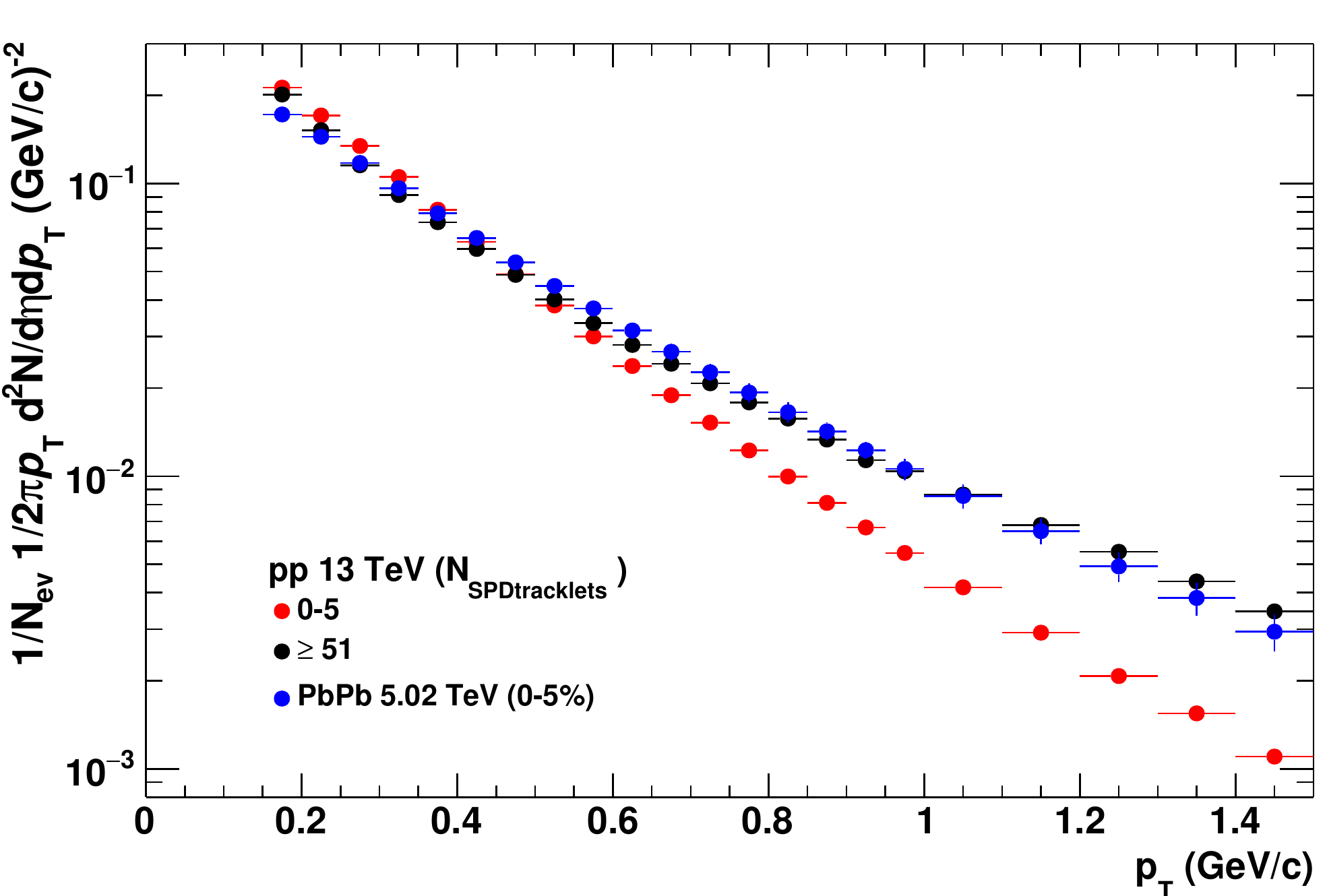}
\vspace*{-0.2cm}
\caption{Invariant transverse momentum distribution of charged particles  from ALICE experiment in ${\it pp}$ collisions at $\sqrt s =$ 13 TeV for two different multiplicity cuts $N_{SPDtracklets} \ge 51 $ (purple solid circle) and $N_{SPDtracklets} < 5$  (red solid circle) \cite{alicepp}. Pb-Pb collision spectra at $\sqrt {s_{NN}} =$ 5.02 TeV for 0-5$\%$ centrality is shown as blue solid circle \cite{alicepb}. }
\label{ptspectra}
\end{figure}

 In the thermodynamic limit, one obtains an analytic expression \cite{pajares1,pajares2}
\begin{equation}
\langle \frac {n S_{1}}{S_{n}} \rangle = \frac {\xi}{1-e^{-\xi}}\equiv \frac {1}{F(\xi)^2}
\end{equation}
where $F\xi)$ is the color suppression factor and $\xi = \frac {N_{s} S_{1}}{S_{N}}$ is the percolation density parameter.
 
Equation~\ref{mu1pt1} can be written as $\mu_{n}=n F(\xi)\mu_{1}$ and 
$\langle p_{T}^{2}\rangle_{n} ={\langle p_{T}^{2} \rangle_{1}}/F(\xi)$.  
The critical cluster which spans $S_{N}$, appears for $\xi_{c} \ge$ 1.2 \cite{inch,satzbook}. 

It is worth noting that CSPM is a saturation model, similar to the Color Glass Condensate (CGC), where $ {\langle p_{T}^{2} \rangle_{1}}/F(\xi)$ plays the same role as the saturation momentum scale $Q_{s}^{2}$ in the CGC model \cite{cgc,perx}. Saturation results from the overcrowding in impact parameter of low $x$ partons of boosted hadrons of nucleus, leading to the appearance of a scale, $Q_{s}^{2}$. This is the basic idea of CGC.  For example, the particle density in CSPM is given by
\begin{equation}
\frac {dn}{dy} \sim (1-e^{-\xi})^{1/2} N_{part}.
\end{equation}
In CGC particle density is related to  the coupling constant $\alpha_{s}(Q_{s}^{2})$,
\begin{equation}
 \frac {dn}{dy} \sim \frac {1}{\alpha_{s}(Q_{s}^{2})} N_{part}.
\end{equation}
While in both cases particle density increases with the number of participants $N_{part}$ \cite{perx}. The correction to the $N_{part}$ scaling in the CGC is due to the occupation number given by $1/\alpha_{s}$ which give rise to a $log(N_{part})$ dependence. In the CSPM, the correction is given by the factor\\ $(1-exp(-\xi))^{1/2}$, which is also a measure of the occupation, indeed is the fraction of the collision area occupied by strings.
\begin{table*}
 \centering
\caption{Event multiplicity classes based on the number of tracklets ($N_{SPDtracklets}$) within $|\eta| <$ 0.8 for $pp$ collisions. For V0M it covers the region 2.8 $< \eta < $ 5.1 and -3.7 $ < \eta < -1.7$. In both cases $ < dN_{ch}/d\eta > $ is given in the region $|\eta| < $ 0.8 \cite{alicepp,alicepp7,alicepb,alicexe}.}
\vspace*{0.5cm}
\begin{tabular}{|c |c |c |c |c |c|}\hline
  System  & ${\it pp}$ 13 TeV  & ${\it pp}$ 13 TeV & ${\it pp}$ 5.02 TeV  & ${\it pp}$ 5.02 TeV & ${\it pp}$ 7 TeV \\ \hline
   & $N_{SPDtracklets}$  & V0M & $N_{SPDtracklets}$ & V0M & V0M \\ \hline
  Multiplicity class & $<dN_{ch}/d\eta>$ & $<dN_{ch}/d\eta>$ & $<dN_{ch}/d\eta>$ & $<dN_{ch}/d\eta>$ & $<dN_{ch}/d\eta>$  \\ \hline
 I    & 54.1 & 26.6 & -  & 19.2  & 21.3 \\ 
  II    & 44.6 & 20.5 & 34.6 & 15.1 & 16.5\\
  III    & 38.9 & 16.7 & 29.9 & 12.4 & 13.5\\
  IV    & 34.1 & 14.3 & 26.2 & 10.7 & 11.5 \\
  V    & 29.3 & 12.6 & 22.4 & 9.47 & 10.1\\
  VI    & 24.5 & 10.6 & 18.5 & 8.04 & 8.45 \\
  VII    & 19.5 & 8.46 & 14.6 & 6.56 & 6.72\\
  VIII    & 14.4 & 6.82 & 10.6 & 5.39  & 5.40  \\
  IX    & 9.03 & 4.94 & 6.58 & 4.05 & 3.90\\
  X     & 2.91 & 2.54 & 2.21 & 2.07 & 2.26 \\\hline
\end{tabular}
\label{table1}
\end{table*}
\section{Determination of color suppression factor $F(\xi)$}
In the present work we have extracted the color suppression factor $F(\xi)$ in high multiplicity events in ${\it pp}$ collisions using ALICE data from the transverse momentum spectra of charged particles at $\sqrt s$ =  5.02 and 13 TeV \cite{alicepp}. ALICE has obtained the transverse momentum distribution for two multiplicity estimators which cover different pseudorapidity regions.The estimators are based on either the total charged deposited in the forward detector (covering the  pseudorapidity regions 2.8 $<\eta<$ 5.1 and -3.7 $<\eta<$ -1.7) V0M or on the number of tracks in the pseudorapidity region $|\eta|<$ 0.8 $N_{SPDtracklets}$. Table~\ref{table1} shows the event multiplicity classes based on V0M and
$N_{SPDtracklets}$ for ${\it pp}$ collisions at $\sqrt s= $ 5.02 and 13 TeV \cite{alicepp}.
Table~\ref{table1} also shows the various multiplicity classes based on V0M estimator for ${\it pp}$ at $\sqrt s =$ 7 TeV \cite{alicepp7}.

Figure~\ref{ptspectra} shows invariant transverse momentum distribution of charged particles for two multiplicity cuts at $\sqrt s=$ 13 TeV in $pp$ collisions and Pb-Pb at $\sqrt {s_{NN}}=$ 5.02 TeV for 0-5 $\%$ centrality.  The spectra become harder for higher multiplicity cuts. This is due to the fact that high string density color sources are created in the higher multiplicity events.
To evaluate the initial value of $F(\xi)$ from data in high multiplicity events in $pp$ collisions, a parameterization of the experimental data of $p_{T}$ distribution in low energy ${\it pp}$ collisions at $\sqrt s=$ 200 GeV  was used \cite{eos}.  The charged particle spectrum is described by a power law \cite{review15}

\begin{equation}
  d^{2}N_{c}/dp_{T}^{2} = a/(p_{0}+p_{T})^{\alpha},
  \label{spectra1}
\end{equation}
where $a$ is the normalization factor, $p_{0}$ and $\alpha$ are fitting parameters with $p_{0}$= 1.98 and  $\alpha$ = 12.87 \cite{eos}. This parameterization is used in high multiplicity ${\it pp}$ collisions to take into account the interactions of the strings \cite{review15}. At low $p_{T}$ the form of function is
approximately exponential with inverse slope parameter $p_{0}/\alpha$ = 154 MeV.  
\begin{figure}
\centering
\vspace*{-0.5cm}
\includegraphics[width=0.6\textwidth,height=3.0in]{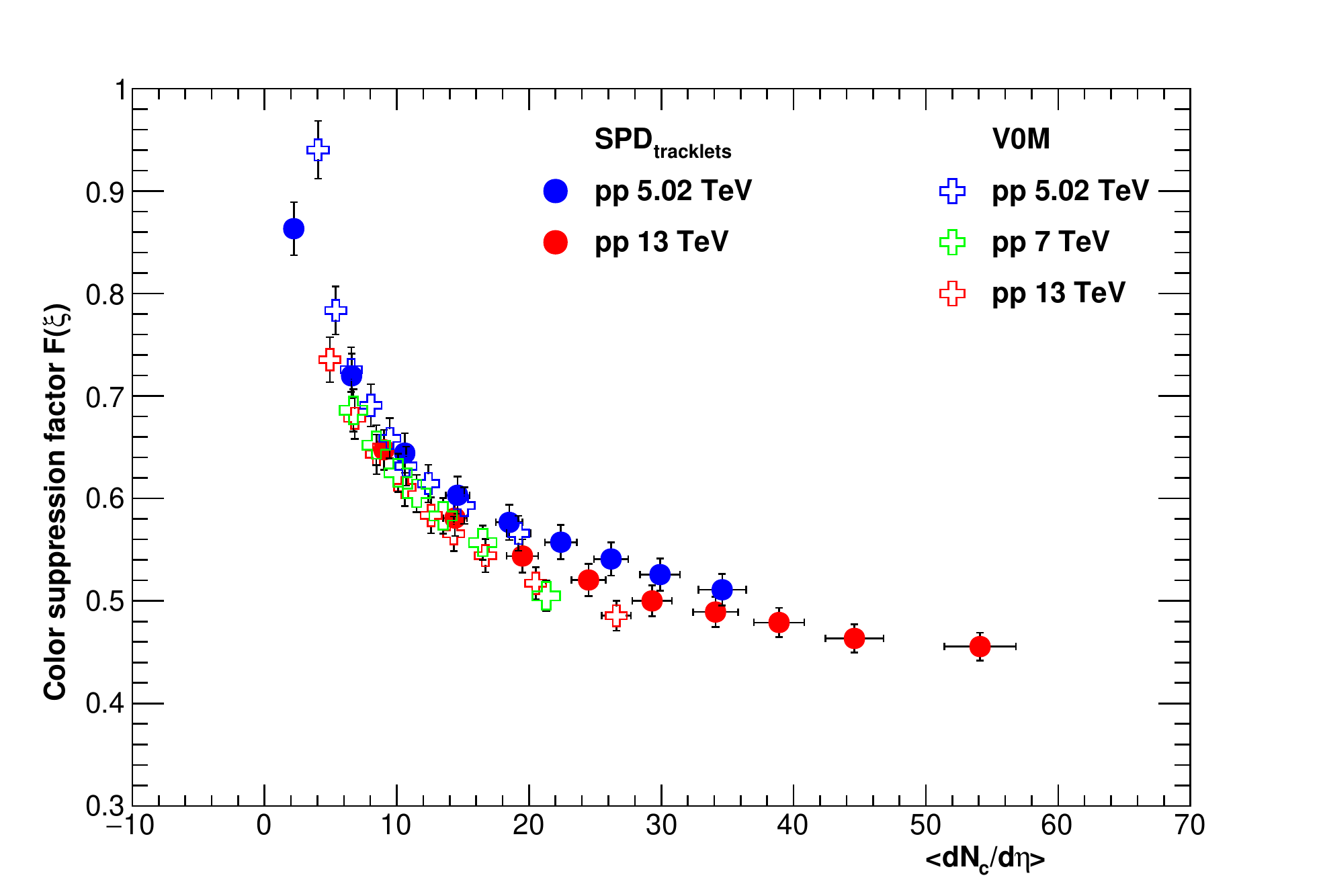}
\vspace*{-0.5cm}
\caption{ Color Suppression Factor $F(\xi)$ in ${\it pp}$ collisions vs  $<dN_{ch}/d\eta>$  for $SPD_{tracklets}$ and V0M event multiplicity classes. For $pp$ collisions at $\sqrt s =$ 7 TeV only V0M event multiplicity classes are available.   $<dN_{ch}/d\eta>$  is the charged particle multiplicity covering the kinematic range $|\eta| < 0.8$ and transverse momentum  $p_{T}$ = 0.15 - 20 GeV/c \cite{alicepp}.}
\label{fxivaluepp7}
\end{figure}
The color suppression factor $F(\xi)$ encodes the effects of the interaction among strings once they overlap. The parameter $p_{0}$ in  Eq.~(\ref{spectra1}) is for independent strings and gets modified
\begin{equation}
  p_{0} \rightarrow p_{0} \left(\frac {\langle nS_{1}/S_{n} \rangle_{pp}^{mult}}{\langle nS_{1}/S_{n} \rangle_{pp}}\right)^{1/4},
  \label{p01}
\end{equation}

\begin{equation} 
  \frac{d^{2}N_{c}}{dp_{T}^{2}} = \frac{a}{(p_{0} \sqrt {F(\xi)_{pp}/F(\xi)_{pp}^{mult}}+{p_{T}})^{\alpha}},
  \label{spectra2} 
\end{equation}
where $F(\xi)_{pp}^{mult}$ is the multiplicity dependent color suppression factor.  In $pp$ collisions $F(\xi)_{pp} \sim$ 1 at low energies due to the low overlap probability. Equation~(\ref{spectra2}) can be written as
\begin{equation} 
  \frac{d^{2}N_{c}}{dp_{T}^{2}} = \frac{a}{(p_{0} \times p_{1}+{p_{T}})^{\alpha}}, \hspace{0.5cm}
  p_{1} = \frac{1}{\sqrt{F(\xi)}}
  \label{spectra3} 
\end{equation}
%
The spectra were fitted using  Eq.~(\ref{spectra3}) in the softer sector with $p_{T}$ in the range 0.2-1.5 GeV/c. The average value of $F(\xi)$ is obtained from varying the fitting range $p_{T}$ = 0.2-1.0,1.2,1.5 GeV/c, $p_{T}$ = 0.25-1.0,1.2,1.5 GeV/c, and  $p_{T}$ = 0.3-1.0,1.2,1.5 GeV/c.

In the thermodynamic limit the color suppression factor $F(\xi)$ is related to the percolation density parameter $\xi $ \cite{review15}
\begin{equation}
   F(\xi) = \sqrt {\frac {1-e^{-\xi}}{\xi}}.
  \label{xi}
\end{equation}
%
%

\begin{figure}[!h]
  \begin{minipage}{0.55\columnwidth}
  \centering
 \includegraphics[width=\textwidth]{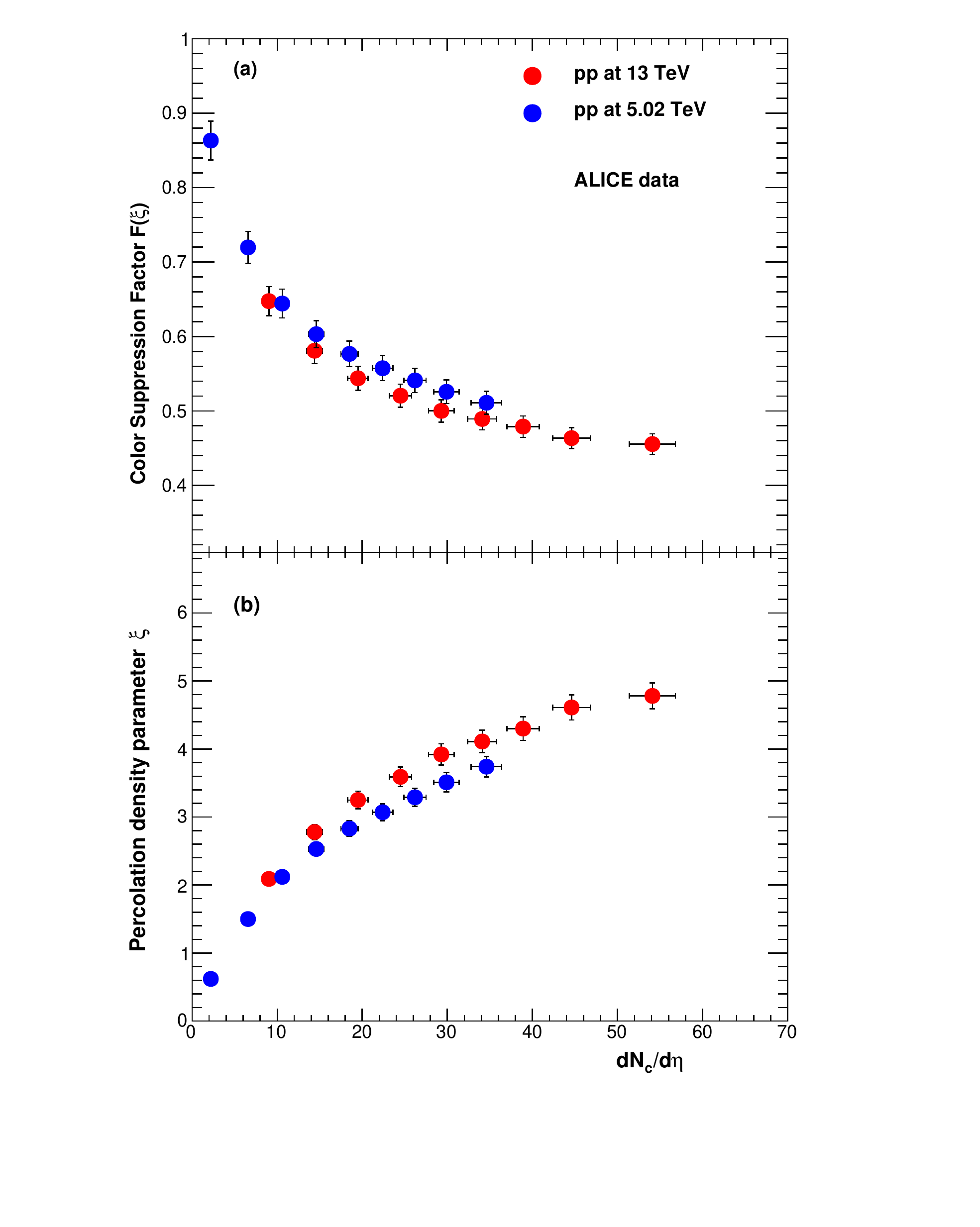} 
\caption{ (a) Color suppression factor $F(\xi)$ and (b)  String density $\xi$ in ${\it pp}$ collisions vs $dN_{ch}/d\eta$. $dN_{ch}/d\eta$ is the charged particle multiplicity covering the kinematic range $|\eta| < 0.8$ and $p_{T}$ =  0.15 - 20 GeV/c \cite{alicepp}. }
\label{fxivaluepp}
\end{minipage}
\hspace{0.5cm}
 \begin{minipage}{0.55\columnwidth}
\centering        
\includegraphics[width=\textwidth]{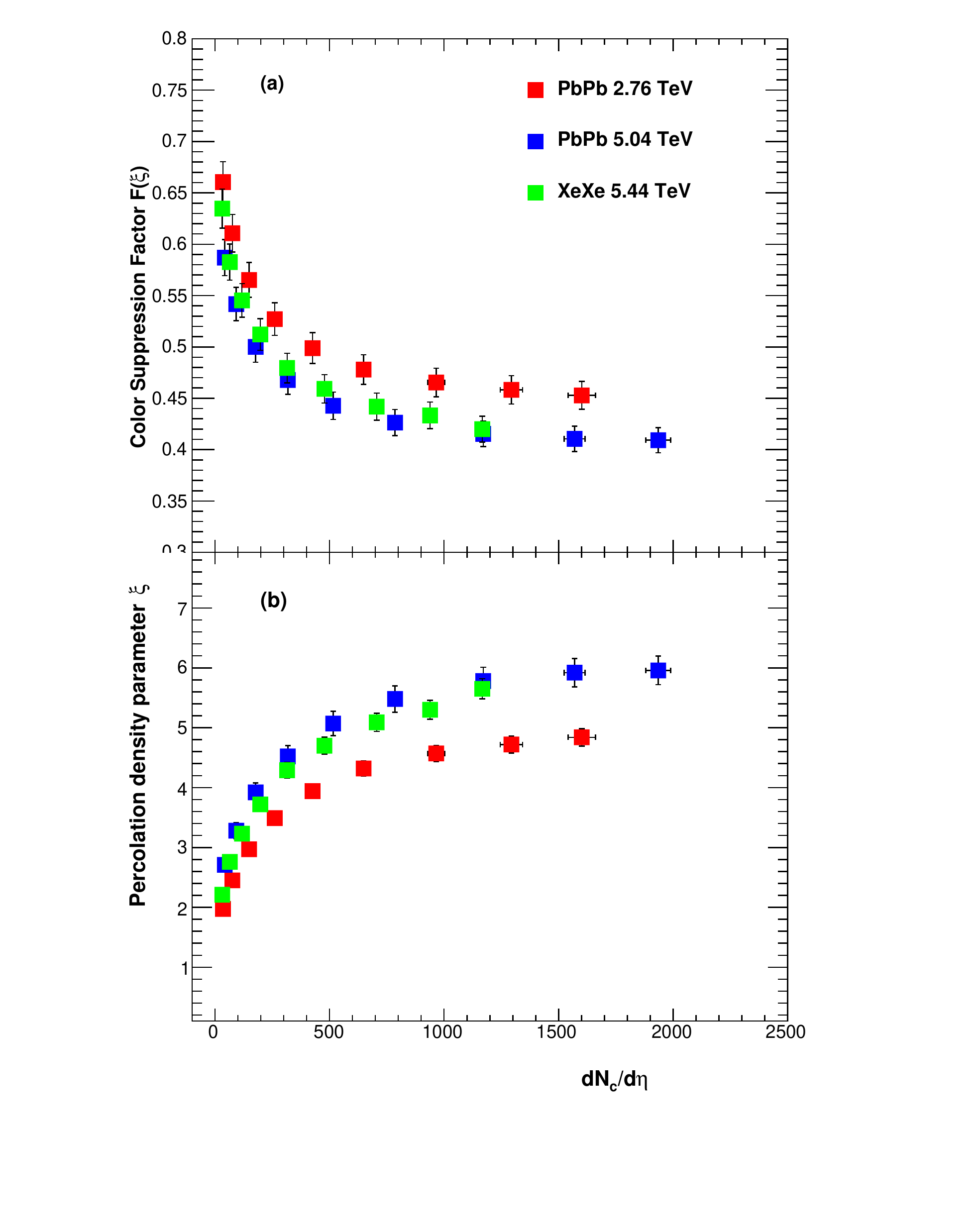}
\caption{(a) Color suppression factor $F(\xi)$ and (b) String density $\xi$ in Pb-Pb, and Xe-Xe collisions vs $dN_{ch}/d\eta$.  $dN_{ch}/d\eta$ is the charged particle multiplicity covering the kinematic range $|\eta| < 0.8$ and  $p_{T}$ = 0.15 - 20 GeV/c \cite{alicepb,alicexe}.}   
\label{fxivalueAA}
\end{minipage}
\end{figure}

Figure~\ref{fxivaluepp7} shows the extracted value of $F(\xi)$ as a function of $<dN_{ch}/d\eta>$ ($N_{tracks}$) from ALICE experiment for $\sqrt {s}$ = 5.02 and 13 TeV using both estimator $SPD_{tracklets}$ and V0M. In case of $\sqrt {s}$ = 7 TeV only V0M estimator results are shown. $N_{tracks}$ is the charged particle multiplicity covering the pseudo-rapidity range $|\eta|<$ 0.8 and  $p_{T}$ = 0.15 - 20 GeV/c \cite{alicepp}. It is observed that  for fixed average charged particle multiplicity $F(\xi)$ has similar values for all  energies. Since $SPD_{tracklets}$ covers higher multiplicity further studies are shown only with this estimator.
Figures~\ref{fxivaluepp} (a) and (b) show $F(\xi)$ and $\xi$ as a function of
$<dN_{ch}/d\eta>$.

\begin{figure}
\centering  
\includegraphics[width=0.50\textwidth,height=2.5in]{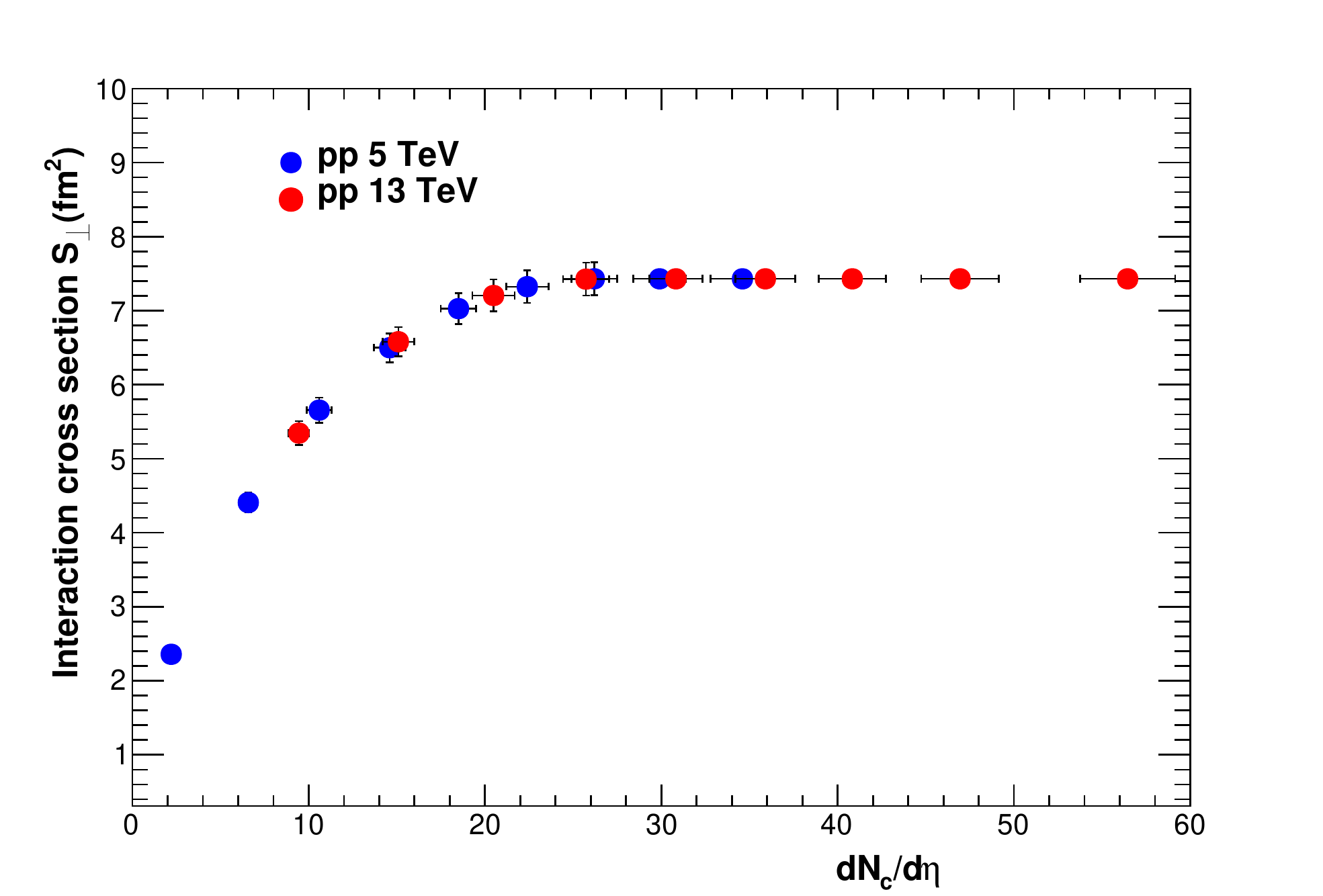}
\vspace*{-0.2cm}
\caption{Interaction cross section $S_{\perp}$ vs $dN_{ch}/d\eta$. $S_{\perp}$ is
  obtained using IP-Glasma model \cite{cross}.}
\label{cross}
\end{figure}

In case of $A-A$ collisions $ {\langle nS_{1}/S_{n} \rangle_{pp}^{mult}}$ 
in Eqs.~(\ref{p01}) and (\ref{spectra2} ) is replaced by ${\langle nS_{1}/S_{n} \rangle_{AA}}$ and $F(\xi)_{AA}$, respectively. For Pb-Pb at $\sqrt {s_{NN}} = $ 2.76 and 5.04 TeV and Xe-Xe at $\sqrt {s_{NN}} = $ 5.44 TeV $F(\xi)$ has been extracted from the spectra at various centralities \cite{alicepb,alicexe}.
Figures~\ref{fxivalueAA} (a) and (b) show $F(\xi)$ and $\xi$ respectively for Xe-Xe and Pb-Pb collisions.
To compare $pp$ with the heavy ions results, we need to normalize $N_{tracks}$
with the transverse area $S_{\perp}$ in ${\it pp}$ and $A-A$ collisions. 	
For ${\it pp}$ collisions $S_{\perp} = \pi R^{2}_{pp}$   has been computed  in the IP-Glasma model, where $R_{pp}$ is the interaction radius \cite{cross1,cross} . This is based on an impact parameter description of $ {\it pp} $ collisions, combined with an underlying description of particle production based on the theory of Color Glass Condensate \cite{cross}.  The interaction radius $R_{pp}$ is approximately a linear function of the charged particle multiplicity. The interaction radius depends on the energy density of the Yang-Mills field and can vary by a factor of $\sim $2 \cite{cross1}.
Here we shall use the parametrization of \cite{cross} and is dependent on the gluon multiplicity 

\begin{equation}
  R_{pp} = f_{pp}(dN_{g}/dy)^{1/3}, 
\label{cross2}
\end{equation}
$f_{pp}= (0.387+0.0335x+0.274x^{2}-0.0542x^{3})$ for $x < 3.4$ and
$f_{pp}$ = 1.538 for $x \geq 3.4$,\\
where $x$ = $(dN_{g}/dy)^{1/3}$. The gluon multiplicity $dN_{g}/dy$ is related to the number of tracks seen in the CMS experiment:
\begin{equation}
dN_{g}/dy \approx (3/2) \frac {1}{\Delta\eta} N_{track}
\label{cross3}
\end{equation}
where $\Delta\eta \sim $ 4.8 units of pseudorapidity.
The interaction cross section $S_{\perp}$  for ${\it pp}$ collisions at $\sqrt {s}$ = 5.02 and 13 TeV from ALICE \cite{alicepp} is shown in Fig.~\ref{cross}. $S_{\perp}$ increases with the multiplicity and for very high multiplicities it is approximately constant.
In the case of A-A collisions the nuclear overlap area was obtained using the
Glauber model. \cite{glauber}.

\begin{figure}[thbp]
\centering        
\vspace*{-0.2cm}
\includegraphics[width=0.65\textwidth,height=3.5in]{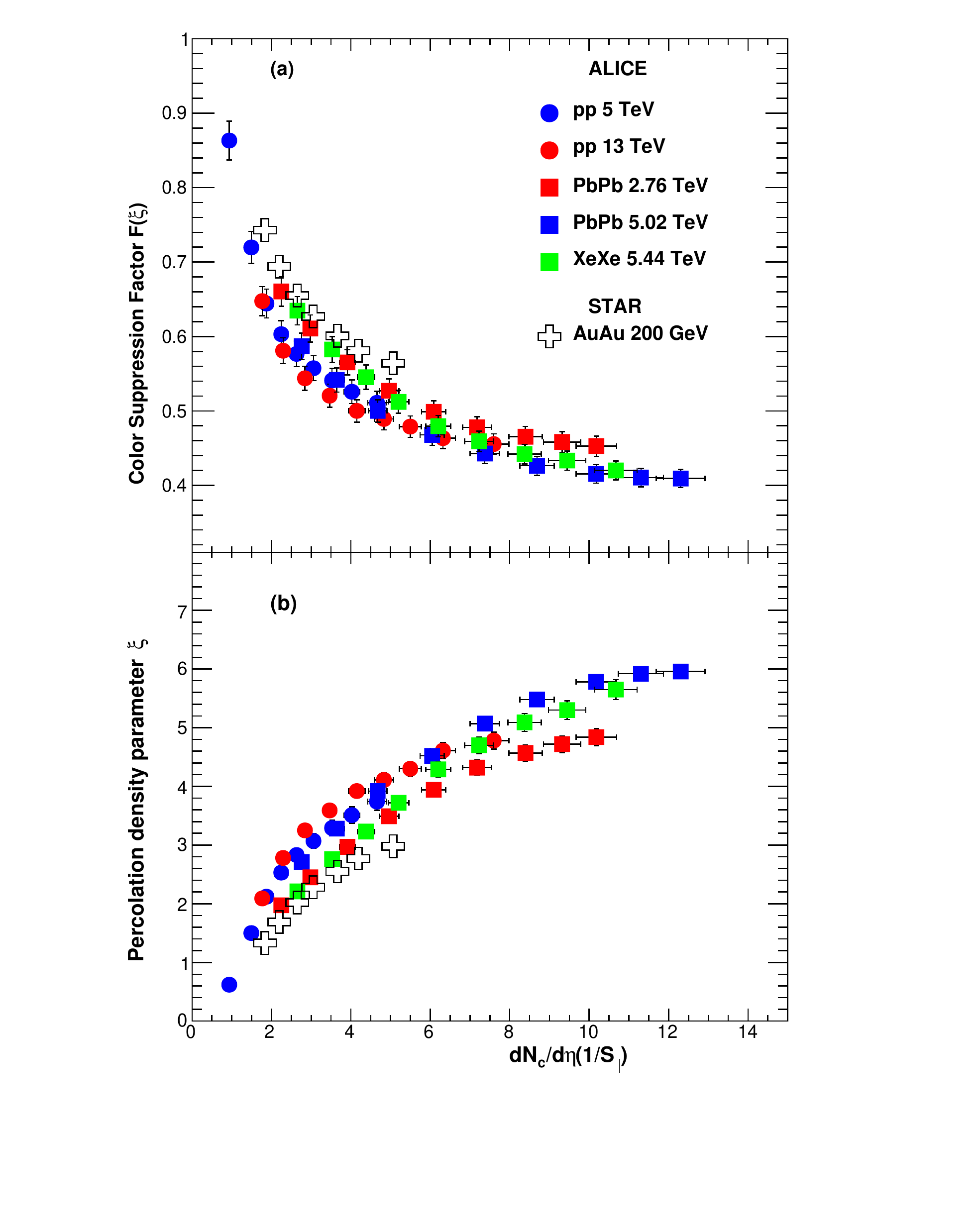}
\vspace*{-1.0cm}
\caption{ (a) Color Suppression Factor $F(\xi)$ in ${\it pp}$, Pb-Pb, and Xe-Xe collisions vs $dN_{ch}/d\eta$ scaled by the transverse area $S_{\perp}$. For ${\it pp}$ collisions $S_{\perp}$ is multiplicity dependent as obtained from IP-Glasma model \cite{cross}. In case of Pb-Pb and Xe-Xe collisions the nuclear overlap area was obtained using the Glauber model \cite{glauber}. (b) String density parameter $\xi$. Results are also shown for Au-Au collisions at  $\sqrt {s_{NN}}$ = 200 GeV \cite{eos}. }
\label{fxintrackarea}
\end{figure}

Figure~\ref{fxintrackarea}(a) shows  $F(\xi)$ as a function of $dN_{ch}/d\eta$ scaled by transverse area $S_{\perp}$ for ${\it pp}$, Pb-Pb, and Xe-Xe collisions. Percolation density parameter $\xi$ is shown in Fig.~\ref{fxintrackarea}(b). Results are also shown for Au-Au collisions at $\sqrt {s_{NN}}$ = 200 GeV
\cite{eos}.  A universal scaling behavior is observed in hadron-hadron and nucleus-nucleus collisions.
%
\section{Connection between $F(\xi)$ and temperature}
The connection between $F(\xi)$ and the temperature $T(\xi)$ involves the Schwinger mechanism (SM) for particle production. 
The Schwinger distribution for massless particles is expressed in terms of $p_{T}^{2}$ \cite{schw,wong}
\begin{equation}
dn/d{p_{T}^{2}} \sim \exp (-\pi p_{T}^{2}/x^{2})
\label{swing}
\end{equation}
where the average value of the string tension is  $\langle x^{2} \rangle$. The tension of the macroscopic cluster fluctuates around its mean value because the chromo-electric field is not constant.
The origin of the string fluctuation is related to the stochastic picture of 
the QCD vacuum. Since the average value of the color field strength must 
vanish, it cannot be constant but changes randomly from point to point \cite{bialas,dosch}. Such fluctuations lead to a Gaussian distribution of the string tension
\begin{equation}
  \frac{dn}{dp_{T}^{2}} \sim \sqrt {\frac{2}{<x^{2}>}}\int_{0}^{\infty}dx \exp (-\frac{x^{2}}
  {2<x^{2}>}) \exp (-\pi \frac{p_{T}^{2}}{x^{2}})
 \label{swingx} 
\end{equation}
 which gives rise to  thermal distribution \cite{bialas}
\begin{equation}
\frac{dn}{dp_{T}^{2}} \sim \exp \left (-p_{T} \sqrt {\frac {2\pi}{\langle x^{2} \rangle}} \right ) ,
\label{bia}
\end{equation}
with $\langle x^{2} \rangle$ = $\pi \langle p_{T}^{2} \rangle_{1}/F(\xi)$. 
The temperature is expressed as \cite{eos,pajares3}  
\begin{equation}
T(\xi) =  {\sqrt {\frac {\langle p_{T}^{2}\rangle_{1}}{ 2 F(\xi)}}}.
\label{temp}
\end{equation} 
%

The string percolation density parameter $\xi$ which characterizes the percolation clusters measures the initial temperature of the system. Since this cluster 
covers most of the interaction area, the temperature becomes a global 
temperature determined by the string density.

We will adopt the point of view that the universal hadronization temperature is a good measure of the upper end of the cross over phase transition temperature $T_{h}$ \cite{bec1}. The single string average transverse momentum  ${\langle p_{t}^{2}\rangle_{1}}$ is calculated at $\xi_{c}$ = 1.2 with the  universal hadronization temperature $T_{h}$= 167.7 $\pm$ 2.6 MeV~\cite{bec1}. This gives \mbox{$ \sqrt {\langle {p_{t}^{2}} \rangle _{1}}$  =  207.2 $\pm$ 3.3 MeV}. 
In this way at $\xi_{c}$ = 1.2 the connectivity percolation transition at $T(\xi_{c})$ models the thermal deconfinement transition. The~temperature obtained using Eq.~ (\ref{temp}) was $\sim$193.6 MeV for Au+Au collisions at \mbox{$\sqrt{s_{NN}}$ = 200 GeV} in reasonable agreement with $T_{eff}$ = 221 $\pm$ $19^{stat} \pm 19^{sys}$ MeV from the enhanced direct photon experiment measured by the PHENIX Collaboration~\cite{phenix}.

The direct photon measurements from ALICE for Pb-Pb collisions at
$\sqrt{s_{NN}}$ = 2.76 TeV in the transverse momentum range 0.9 $< p_{T} < $ 2.1 GeV/c for the 0-20 $\%$ class gives the inverse slope parameter $T_{eff}$  = 297 $\pm 12^{stat} \pm 41^{sys}$ MeV \cite{alicetemp}. In the present analysis the CSPM temperature for 0--5 $\%$ centrality is T = 230 $\pm 7$ MeV. In a recent work using the state-of-the-art hydrodynamics simulations an effective temperature $T_{eff}$ = 222$\pm$9 MeV was obtained in central PbPb collisions at $\sqrt {s_{NN}}$ = 5.02 TeV \cite{gardim}.

\begin{figure}[thbp]
\centering        
\vspace*{1.0cm}
\includegraphics[width=0.6\textwidth,height=3.0in]{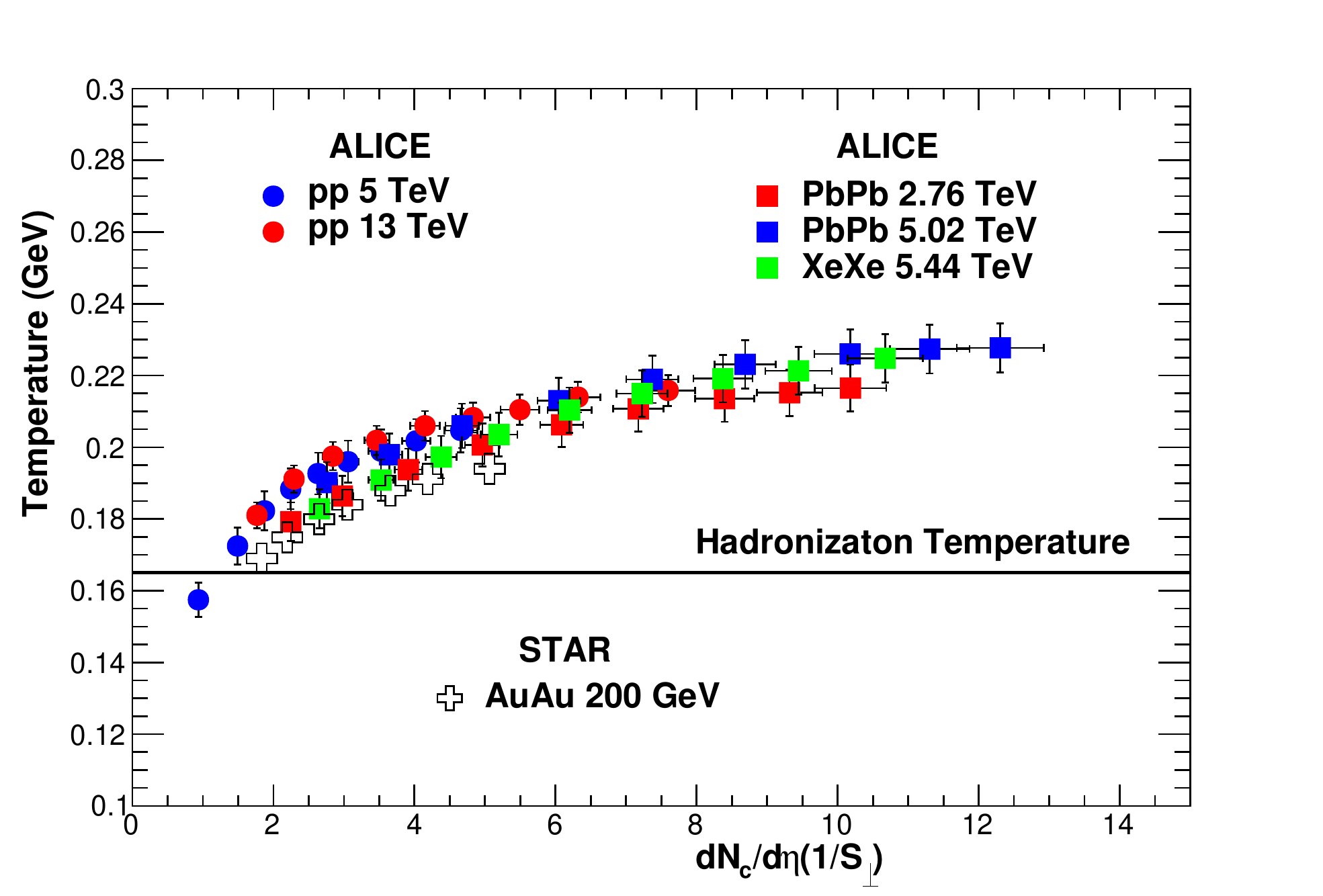}
\vspace*{-0.2cm}
\caption{ Temperature vs  $dN_{ch}/d\eta$ scaled by  $S_{\perp}$ from ${\it pp}$, Pb-Pb, and Xe-Xe collisions. The horizontal  line  at temperature T $\sim 165 $ MeV is the universal hadronization temperature \cite{bec1}. } 
\label{tempfig}
\end{figure}

\begin{figure}[thbp]
\centering        
\vspace*{0.5cm}
\includegraphics[width=0.65\textwidth,height=3.5in]{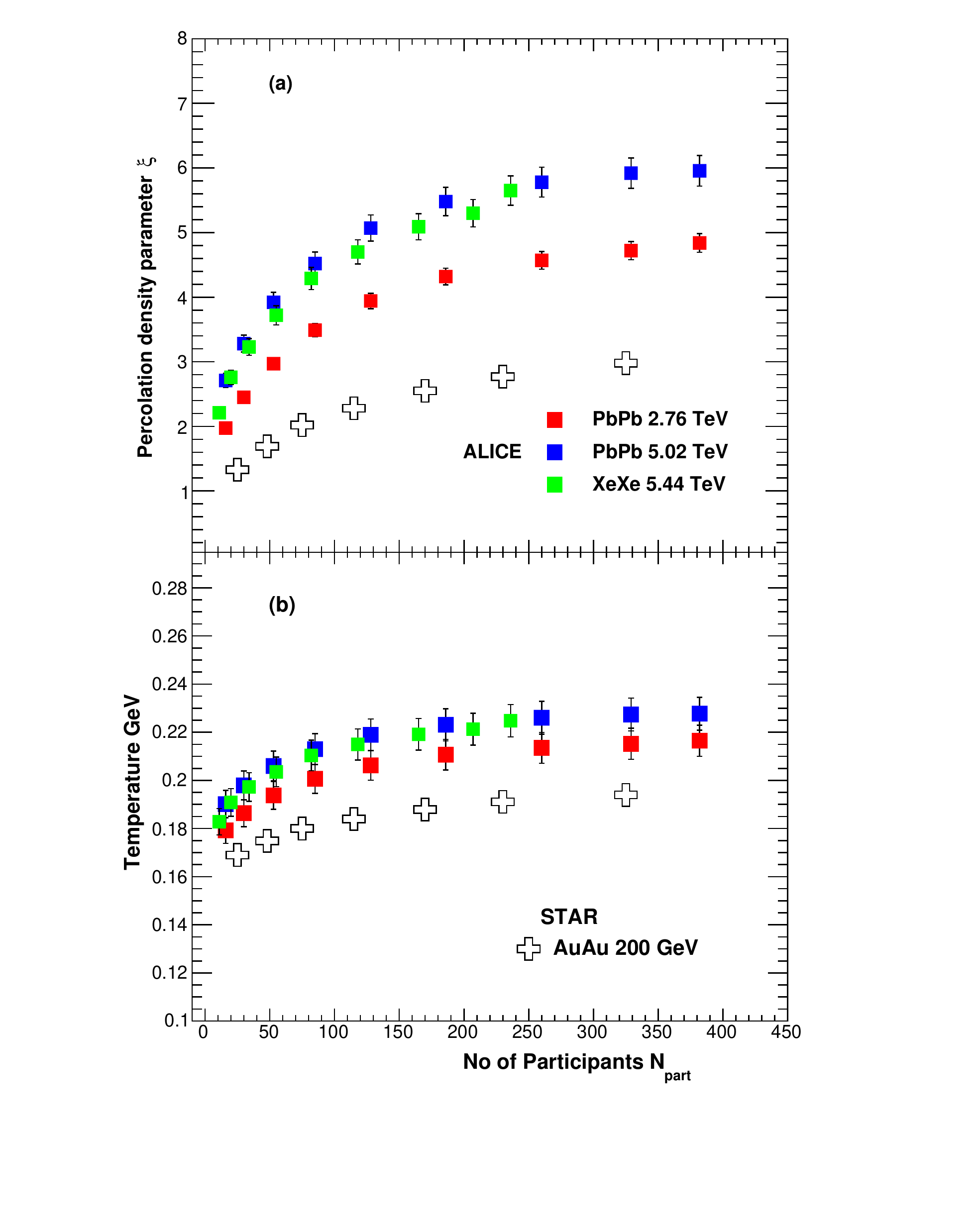}
\vspace*{-1.0cm}
\caption{(a) Percolation density parameter $\xi$, and (b) Temperature as a function of number of participants $N_{part}$ in Pb-Pb and Xe-Xe collisions along with Au-Au at  $\sqrt {s_{NN}}$ = 200 GeV \cite{eos}.}
\label{npart}
\end{figure}

Figure \ref{tempfig} shows the temperature as a function of $dN_{c}/d\eta$ scaled by the interaction area  for $pp$, Xe-Xe, and Pb-Pb collisions at LHC energies. Temperatures from both hadron-hadron and nucleus-nucleus  collisions fall on a universal curve when the multiplicity is scaled by the transverse interaction area. The horizontal line at temperature T $\sim $ 167.7 MeV is the universal hadronization temperature obtained from the systematic comparison of the statistical thermal model parametrization of hadron abundances measured in high energy  $e^{+}e^{-}$, $pp$, and $A-A$ collisions \cite{bec1}. In Fig.~\ref{tempfig} for ${\it pp}$ collisions at $\sqrt {s}$ = 5.02 and 13 TeV higher multiplicity cuts show temperatures above the hadronization temperature and similar to those observed in Au-Au collisions at $\sqrt {s_{NN}}$ = 200 GeV \cite{eos}.
 It is observed that the temperature rises slowly at higher $dN_{c}/d\eta$ values. This behavior is related to the overlap area of strings, which is given by
 ($1-e^{-\xi}$). Above $\xi \sim $ 4 there is complete overlap and the
 temperature  rises slowly as  $\xi^{1/4}$.
 In Figs.~\ref{npart} (a) and (b) $\xi$ and temperature are shown as a function of number of participants $N_{part}$. It is observed that the string density $\xi$ is collision energy dependent for the same number of participants. 

\section{Thermalization}
In case of $A-A$ collisions it was observed that the average transverse momentum is twice the temperature $<p_{T}> \sim $ 2T \cite{review15}. This shows that the  charged particle transverse momentum spectrum is exponentially distributed, and the inverse slope parameter is the thermalized temperature \cite{review15}. Similar behavior is observed in $pp$ collisions as shown in Fig.~\ref{pt2t} along with Pb-Pb and Xe-Xe results.

In the CSPM each cluster of strings have different tension corresponding to the resulting color field from the sum of the individual color field of the overlapping strings. The distribution of these cluster size is the Gamma distribution
\cite{review15}. Each cluster decays into a shape $exp({-p_{T}^{2}/a})$ which transforms into a thermal distribution assuming a Gaussian for the string fluctuations. The convolution of the Gamma distribution with a thermal distribution gives rise to the
\begin{equation}
  \frac{1}{(1+{p_{T}/n<p_{T}>})^{n}}
    \end{equation}
 with $<p_{T}> = <p_{T}>_{1}/\sqrt{ F(\xi)}$. $<p_{T}>_{1}$ is the mean $<p_{T}>$ of particles produced by a single string. This distribution is similar to the used in Eq.~\ref{spectra2}, where  $p_{0}$ is related to $<p_{T}>_{1}$.
It has been suggested that fast thermalization in ${\it pp}$ and $A-A$ collisions can also occur through the existence of an event horizon caused by a rapid deceleration of the colliding nuclei \cite{casto1,alex}. The thermalization is due to the Hawking-Unruh effect \cite{casto1,hawk,unru}. 
It is well known that the black holes evaporate by quantum pair production and behave as if they have an effective temperature of 
\begin{equation}
T_{H}= \frac {1}{8\pi GM},
\end{equation}
where 1/4GM is the acceleration of gravity at the surface of a black hole of mass M. The rate of pair production in the gravitational background of the black hole can be evaluated by considering the tunneling through the event horizon. Unruh showed that a similar effect arises in a uniformly accelerated frame, where an observer detects the thermal radiation with the  temperature T =a/2,
where $a$ is the acceleration. Similarly, in hadronic interactions the probability to produce states of masses M due to the chromoelectric field E and color charge is given by the Schwinger mechanism \cite{schw}
\begin{equation}
W_{M} \sim \exp (\frac {-\pi M^{2}}{gE})\sim \exp(-M/T),
\end{equation}
which is similar to the Boltzmann weight in a heat bath with an effective temperature
\begin{equation}
T = \frac {a}{2\pi},\, a= \frac {2gE}{M}.
\end{equation}
In CSPM the strong color field inside the large cluster produces deceleration of the primary $q \bar q$ pair which can be seen as a thermal temperature  by means of the Hawking-Unruh effect. This implies that the radiation temperature is determined by the transverse extension of the color flux tube/cluster in terms of the string tension \cite{casto1,alex,casto2}. 
\begin{equation}
T =  \sqrt {\frac {\sigma}{2\pi}}
\label{haw}
\end{equation}
where $\sigma$ is string tension. This string tension referred to the tension of a cluster of strings can be written in terms of the color suppression factor in such a way that Eq.~(\ref{haw}) is same as Eq.~(\ref{temp}). This is not surprising because both the Hawking-Unruh effect and Eq.~(\ref{temp})  are based on the Schwinger mechanism. The string percolation density parameter $\xi$ which characterizes the percolation clusters measures the initial temperature via the color reduction factor F($\xi$). Since the cluster covers most of the interaction area, this temperature becomes a global temperature.
\section {Energy density}
\begin{figure}[!h]
  \begin{minipage}{0.55\columnwidth}
  \centering
      \includegraphics[width=\textwidth]{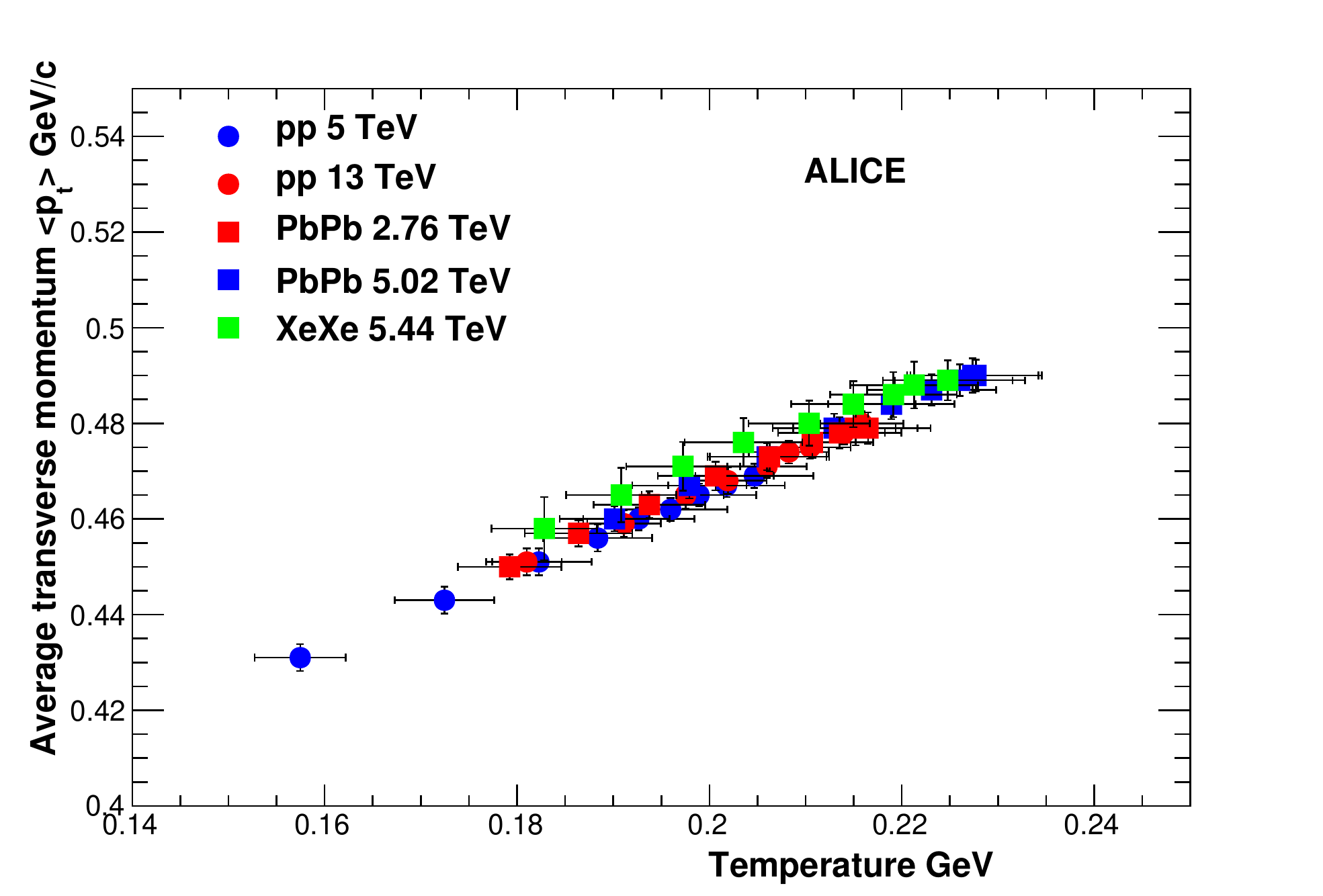}
 \caption{Average transverse momentum vs temperature.} 
\label{pt2t} 
\end{minipage}
\hspace{0.5cm}
\begin{minipage}{0.50\columnwidth}
\centering        
\includegraphics[width=\textwidth]{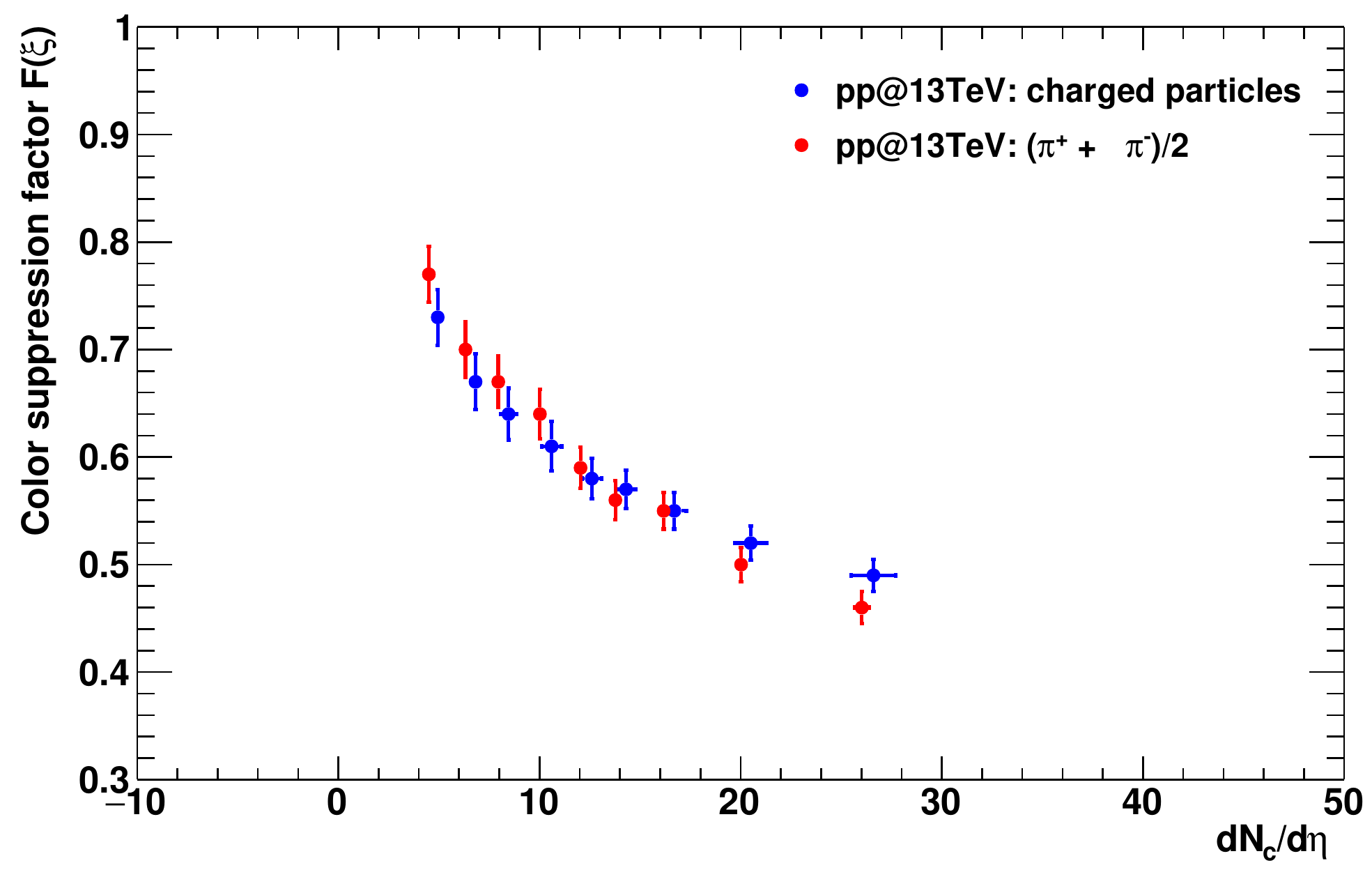}
\caption{Color suppression factor  $F(\xi)$ for charged particles and pions in ${\it pp}$ collisions vs  $<dN_{ch}/d\eta>$  for V0M event multiplicity classes . $<dN_{ch}/d\eta>$ is the charged particle multiplicity covering the kinematic range $|\eta| < 0.8$ and transverse momentum  $p_{T}$ = 0.15 - 20 GeV/c \cite{alicepp}.}
\label{fxipion}
\end{minipage}
\end{figure}

The energy density is obtained using the well known Bjorken boost invariant 1D
hydrodynamics \cite{bjorken}
 
\begin{equation}
\bm \varepsilon = \frac {3}{2}\frac { {\frac {dN_{c}}{dy}}\langle m_{T}\rangle}{S_{n} \bm {\tau_{pro}}},
\label{bjk}
\end{equation}
where $\bm \varepsilon$ is the energy density, $S_{n}$ the nuclear overlap area, $m_{T}$  transverse mass. The quantity $\bm {\tau_{pro}}$, the production time
at which quark-gluon plasma is produced, is given by \cite{wong}

\begin{equation}
\bm {\tau_{pro}} = \frac {2.405\hbar}{\langle m_{T}\rangle}.
\end{equation} 
Above the critical temperature only massless particles are present in CSPM. The average transverse mass $<m_{T}>$ is given by $<m_{T}> = \sqrt{<p_{T}>^{2} + m_{0}^{2}}$, $m_{0}$ being the mass of pion. Since the color suppression factor $F(\xi)$ is same for both charged particles  and pions in the $p_{T}$ range 0.12-1.0 GeV/c,  we have used pion mass to obtain the energy density. Average $p_{T}$ is obtained from the experimental data in the same $p_{T}$ range as used in fitting the spectra. 
Figure~\ref{fxipion} shows the color suppression factor $F(\xi)$ for both charged particles and pion for V0M event multiplicity classes. It is observed that $F(\xi)$ remains same for both pions and charged particle. 
  
In Figure~\ref{enxi2} energy density $\varepsilon$ as a function of string density is shown for $pp$ and $A-A$ collisions at LHC energies. Results for ${\it pp}$ at $\sqrt {s}$ = 13 TeV  and Pb-Pb at $\sqrt {s_{NN}}$ = 2.76 and 5.02 TeV are from our earlier publication \cite{epja21}. Figure~\ref{enxi2} also has results from Au-Au at  $\sqrt {s_{NN}}$ =200 GeV \cite{eos}. 
We observe a slow rise of  $\bm \varepsilon$  for low values of ${\xi}$ followed by a faster rise later. It is found  that $\bm \varepsilon$ is proportional to $\xi$ in the range 
$1.2 < \xi < 5.0 $. Above $\xi \sim $  5 the energy density $\bm \varepsilon$ rises faster compared to $\xi < $ 5.
A possible explanation for the sharp rise in the energy density could be that at such high degree of overlapping the gluons are seen naked without interaction and thus the coherence of the color fields of the overlapping strings is lost and thus recover the independence of the strings. This means that instead of a dependence on $N_{part}$ it would be $N_{part}^{4/3}$.

It is worth mentioning that the CSPM is only used to encode the dependence on the centrality and energy of the temperature T given by the $p_{T}$ spectrum.
As far as this evaluation can be done directly from the $p_{T}$ spectrum independently of the CSPM, our result can be seen independent of the model used.
%

Energy density has been obtained in lattice set up of (2+1)-flavor QCD using
 the highly improved staggered quark (HISQ) action and the tree-level improved gauge action \cite{lattice14,wuppe14}.
Figure \ref{et4} shows dimensionless quantity $\bm \varepsilon/T^{4}$ as a function of temperature both from CSPM and LQCD.
It is observed that CSPM results differ from LQCD results above the temperature of T $\sim$  210 MeV in case of Pb-Pb and Xe-Xe collisions. In $pp$ collisions
CSPM results are lower as compared to LQCD simulations.  
Beyond T$\sim$ 210 MeV the
$\bm \varepsilon/T^{4}$ in CSPM rises much faster and reaches the ideal gas value of  $\bm \varepsilon/T^{4}$ $\sim$ 16 at T $\sim$ 230 MeV. In this region, there is a strong screening due to the large degree of overlapping of the strings, producing a faster approach to the quark gluon gas limit.

There are some uncertainties which can modify the normalization of  Figures~\ref{enxi2} and \ref{et4}, but not there shape. We obtain the temperature using  Eq.~(\ref{temp}) to fit the $p_{T}$ spectrum in the softer region 0.2 $ < p_{T} < 1.5 $ GeV/c as most of the particles are produced in this range. The use of this range is to remain in the thermal range of $p_{T}$ distribution. The broader range in the $p_{T}$ spectrum would modify the obtained mean $p_{T}$ as well as the temperature. In fact this can reduce the value of $\bm \varepsilon/T^{4}$. The uncertainties in energy density and temperature work in the opposite direction in the  evaluation of $\bm \varepsilon/T^{4}$. Both uncertainties do not modify the shape of the curves in Fig. \ref{et4}, including the change of slope of the curve $\sim $ 210 MeV. Our results are consistent with the normalization of  $\bm \varepsilon/T^{4}$ found at lower temperature with the LQCD results.

The interaction of strings spend a time after the collision to form clusters during the expansion of the system and thus the ~Eq.(\ref{temp}) for the temperature should be able to describe the radial flow as it does. The temperatute obtained using ~Eq.(\ref{temp}) can be compared with the blast wave model \cite{bl1,bl2} expressed in terms of kinetic freeze out temperature $T_{0}$ and the mean radial velocity $\beta$. The resulting temperature T is larger than the $T_{0}$ because the color reduction factor is smaller than one. 
It is characterized by initial temperature because this is approximately the temperature of the ensamble of quarks and gluons coming from the fragmentation of strings created by Schwinger mechanism. These quarks and gluons recombined to form the particles whose transverse momentum we fit to the temperature.

  The interaction gives rise to the azimuthal dependence of the produced particle distribution as it is described by the CSPM in agreement with the experimental data \cite{ellp1,ellp2}. However, this interaction is small and does not modify the
$p_{T}$ distribution of the produced particles and consequently the obtained temperature.

We are aware of the event by event and inside event temperature fluctuations. These fluctuations are translated into $p_{T}$ fluctuations which are well described by the CSPM \cite{ellena}. 

\section{CSPM and two temperatures }
In CSPM, we can associate two scales and the corresponding temperatures to the two different transitions. One is the string density required to have a large cluster of strings crossing the surface of the collision. This happens at the critical percolation
density
\begin{equation}
  \frac {N}{S_{\bot}} = \frac {\xi_{c}}{\pi r_{0}^{2}}
\end{equation}
At this critical percolation density a cluster of strings is extended over most of the collisions surface and so the color field. The corresponding temperature to this critical density can be associated to the confined-deconfined transition. 
  On the other hand, as the area covered by the strings 
 is $(1-exp(-\xi))S_{\bot}$, the mean distance between strings $d$ is given by
\begin{equation}
  d = \left [\frac{N}{(1-exp(-\xi))S_{\bot}} \right]^{-1/2}= F(\xi)\sqrt{\pi}r_{0}.
\end{equation}
For small $\xi$ , $d > r_{0}$ for example at the critical percolation density
$\xi_{c} $ = 1.2, d= 1.34$ r_{0}$. This means that overlapping between strings is very peripheral, covering only the edges(corona) of the strings. This corresponds to the hadronization temperature of $\sim$ 166 MeV. 
In order to penetrate the core, the overlapping should be larger in such a way
that $d = 2h < r_{0}$. For reasonable values of $h$ $\sim$ 0.4, 0.5 ($\sim$ half of the ratios of the string), we have $F(\xi) \sim 0.45-0.50$
corresponding to the temperature of $\sim$ 210 - 220 MeV. This is the temperature at which our result starts deviating from LQCD. The strong overlapping of strings penetrating the hard core means that we are deep inside the color cloud surrounding the source in such a way that the source appears undressed.

From the Statistical  point of view of in order to have a liquid, the radial distribution of the constituents of the regarded system should have a characteristic shape which only can be obtained if the constituents have a repulsive hard core \cite{ramirez}. This feature of liquid of quarks and gluons have been obtained in A-A and $pp$ collisions. This can be applied to CSPM as well. This means that strings should have a hard core if we need to obtain such a liquid. Then our second scale may mark the size of the hard core and possible transition from a liquid of Strong interacting quarks and gluons to a free gas.  

\section{ Degrees of freedom DOF}
\begin{figure}[!h]
  \begin{minipage}{0.6\columnwidth}
  \centering
      \includegraphics[width=\textwidth]{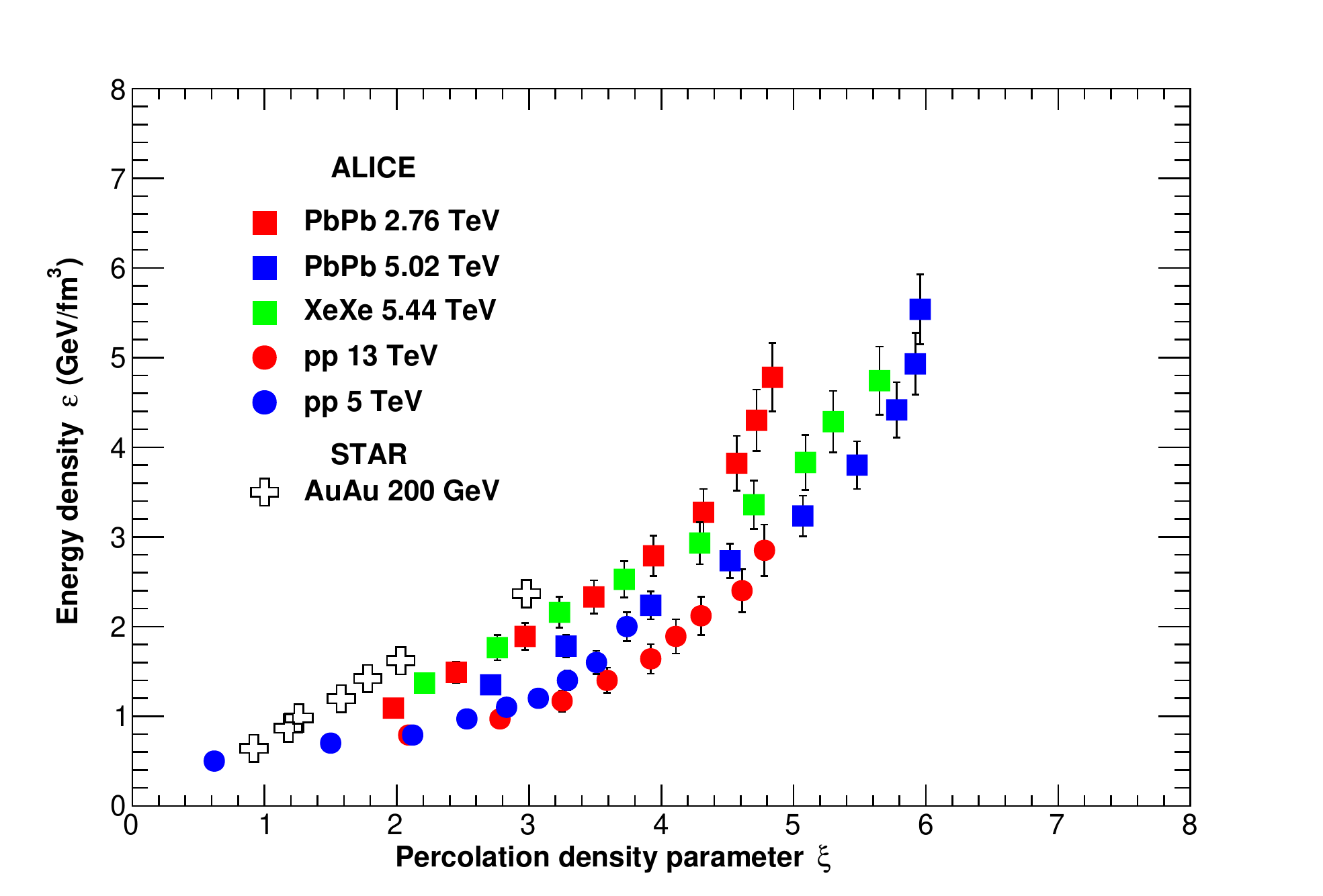}
 \caption{ Energy density ($\epsilon$) as a function 
   of the percolation density parameter ($\xi$) for pp collisions
 at $\sqrt s $ = 5.02 and 13 TeV, Pb-Pb collisions at
 $\sqrt {s_{NN}}$ = 2.76 and 5.02 TeV, and Xe-Xe collisions at $\sqrt {s_{NN}}$ = 5.44 TeV. Results from Au-Au collisions at $\sqrt {s_{NN}}$ = 200 GeV are also shown \cite{eos}.} 
\label{enxi2}
\end{minipage}
\hspace{0.5cm}
\begin{minipage}{0.6\columnwidth}
\centering        
\includegraphics[width=\textwidth]{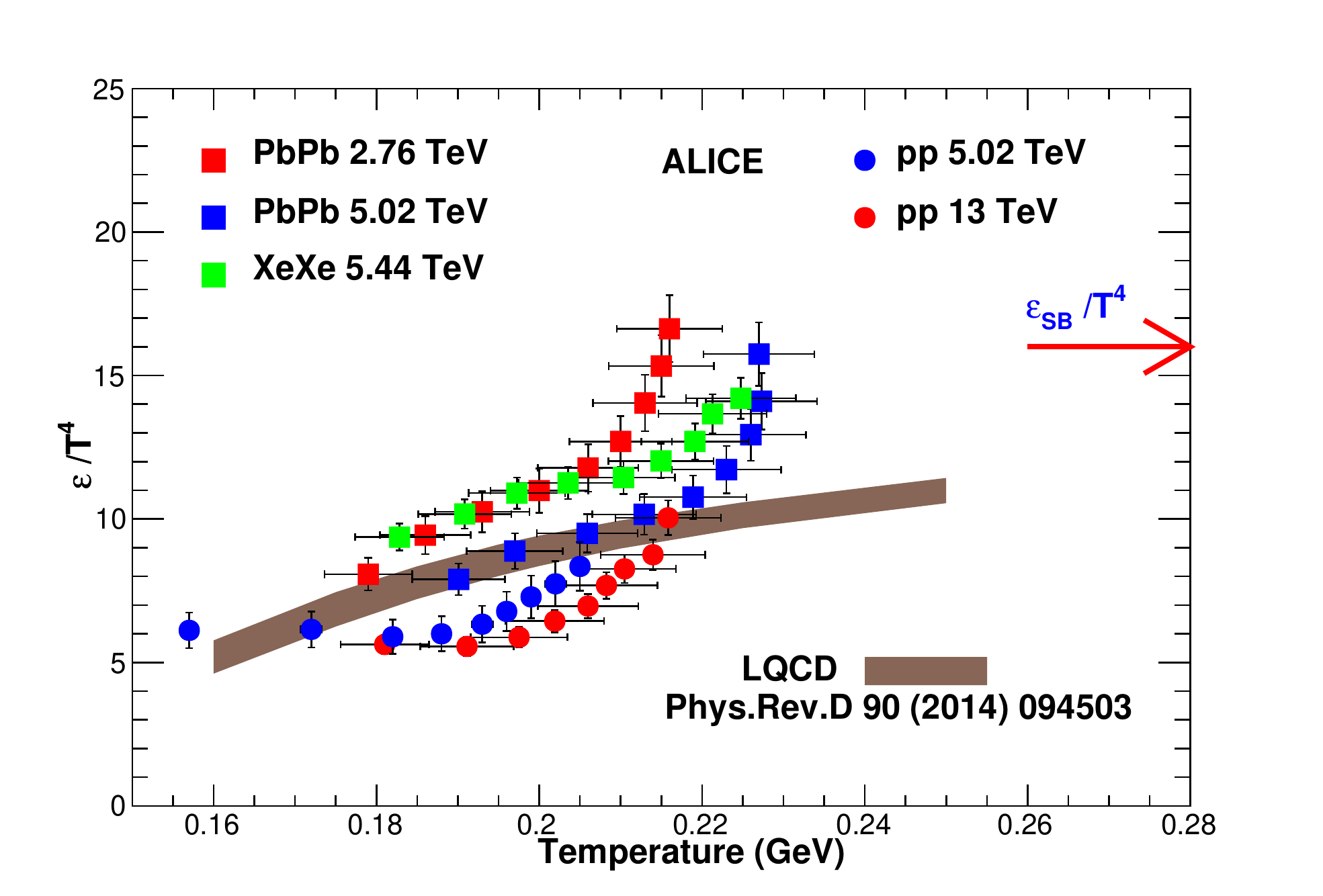}
\caption{ Dimensionless quantity $\bm \varepsilon/T^{4}$ as a function of temperature from CSPM and LQCD calculation from HotQCD Collaboration \cite{lattice14}. }

\label{et4}
\end{minipage}
\end{figure}

In case of a quark-gluon system in thermal equilibrium at a high temperature the quarks and gluons are idealized to be non-interacting and massless and there is no net baryon number. The number of quarks and antiquarks are equal \cite{wong,satzbook}. The energy densities obtained in full QCD with different numbers of quark flavor are given by

\begin{equation}
  \bm \varepsilon/T^{4} \simeq (37/30)\pi^{2} \simeq 12, N_{f}=2.
  \label{nf2}
\end{equation}
\begin{equation}
  \bm \varepsilon/T^{4} \simeq (47.5/30)\pi^{2} \simeq 16, N_{f}=3.
  \label{nf3}
\end{equation}
At T $\sim$ 210 MeV, $\bm \varepsilon/T^{4}$ $\sim$ 11 which corresponds to $\sim$ 33 DOF while at T  $\sim$ 230 MeV there are $\sim$ 47 DOF.
In Fig.~\ref{et4} Stefan-Boltzmann limit is also shown at
$\bm \varepsilon_{SB}/T^{4}$, which corresponds to $\sim$ 48 DOF. 
It is observed that Pb-Pb at $\sqrt {s_{NN}}$ = 2.76  TeV has similar features as seen at 5.02 TeV. For Xe-Xe collisions at $\sqrt {s_{NN}}$ = 5.44 TeV $\sim$ 44 DOF is obtained. 
In ${\it pp}$ collisions at  $\sqrt {s}$ = 13 TeV only $\sim$ 33 DOF are reached. Our results agree with the conclusions obtained studying the trace anomaly in a quasi particle gluonic model \cite{mann1,mann2}. In this model the DOF of the free gluons are also obtained for T $\simeq$ 1.3T$_{c}$ (T$_{c} \approx$ 165 MeV).
\section{Discussion}
The most pressing question to be addressed is the appearance of a transition in CSPM at T $\sim$ 210 MeV above the deconfinement transition.

A new phase in QCD also has been proposed studying the Dirac operator \cite{glozman1}. While confining chromo-electric interaction is distributed among all modes of Dirac operator, the chromo-magnetic interaction is located predominantly in the near zero modes. Above T $\sim$ 155 MeV the near zero modes are suppressed but not the rest of the modes, surviving the chromo-electric interaction which is suppressed at higher temperature \cite{glozman1}.

Recently, it has been pointed out  using lattice simulations that in addition of the standard crossover phase transition at T $\sim$ 155 MeV, the existence of a new infrared phase transition at temperature T, 200 $<T_{IR}<$ 250 MeV. In this phase, asymptotic freedom works and therefore there is no interaction. In between these two temperatures there is possible coexistence of the short and long-distance scales \cite{newphase1,newphase2}, which supports the present observation in our work.

An alternative explanation of results found here is related to the strong magnetic field which is produced in  pp collisions and even stronger in $A-A$  collisions.
The lattice studies of the QCD chiral phase transition with three flavor in a background magnetic field show that chiral condensate and thus the temperature of this phase transition always increases with  magnetic field.
The transition instead of a crossover becomes a first order phase transition. As magnetic field  is higher in Pb-Pb than in $pp$, a higher
temperature is expected for heavy ion collisions \cite {mag2}. This is in agreement with Fig.~\ref{et4}.

 \section{Conclusions}
 We have used the Color String Percolation Model (CSPM) to explore the initial stage in ${\it pp}$, Xe-Xe, and Pb-Pb collisions at LHC energies and determined the thermalized initial temperature of the hot nuclear matter at an initial time $\sim$ 1 $fm/c$. For the first time both the temperature and the energy density of the hot nuclear matter have been obtained from the measured charged particle spectra using ALICE data for ${\it pp}$ collisions at $\sqrt {s}$ = 5.02 and 13 TeV along with Pb-Pb at $\sqrt {s_{NN}}$ = 2.76 and 5.44 TeV and 
 Xe-Xe at $\sqrt {s_{NN}}$ = 5.44 TeV.
 A universal scaling in the  temperature  is obtained
 for both $pp$ and $A-A$ (Fig.~\ref{tempfig}) and are well above the universal hadronization temperature indicating that the matter created is in the deconfined phase. The thermalization in both $pp$ and $A-A$ is reached through the stochastic process ( Hawking-Unruh) rather than the kinetic approach.  
  
 The dimensionless quantity $ \varepsilon/T^{4}$ is evaluated to obtain the number of degrees of freedom (DOF) of the deconfined phase.
We observe two features hitherto not reported: the existence of two temperature ranges in the behavior of the $A-A$ system DOF,
and a clear departure from the LQCD results regarding the maximum number of DOF, which reaches values in agreement with the Stephan Boltzmann limit for an ideal gas of quarks and gluons. In case of $pp$ collisions at $\sqrt {s}$ = 5.02 we reach only $ \varepsilon/T^{4}$ $\sim$ 8 corresponding to $\sim$ 24 DOF, while at $\sqrt {s}$ = 13 TeV $\sim$ 30 DOF is obtained.

It is worth noting the importance of our main result, namely the departure of
 $\varepsilon/T^{4}$ from the LQCD result starting at $T \sim 210-220$ MeV. 
As the energy density is obtained directly from the ALICE data on $dN_{c}/d{\eta}$
 and $p_{T}$ distributions and the temperature is determined by the mean $p_{T}$ which is obtained  directly form the data, the result can be considered as an experimental result. 

It has been argued that QCD could lead to a three-phase structure as a function of the temperature. In such a scenario, color deconfinement would result in a plasma of massive ``dressed'' quarks; the only role of gluons in this state would be to dynamically generate the effective quark mass. The effective DOF in the resulting quark plasma thus are just those of massive quarks.  
At still higher temperature, this gluonic dressing of quarks would then evaporate leading to the QGP \cite{satzbook}.

\section{Acknowledgments}
G. P. work is supported by the grant from CONACYT under the grant A1-S-22917 and CF-2042. 
C. P. thanks the grant Maria de Maeztu Unit of excellence MDM-2016-0682 of Spain, the support of Xunta de Galicia under the project  ED431C 2017 and  project PID 2020-119632GB-IOO of Ministerio de Ciencia e Innovacion of Spain and FEDER.

\end{document}